\documentclass{ws-mplb}

\begin{document}

\markboth{Jian-Jun Dong}{The A-Cycle Problem In XY model with Ring Frustration}

%
\catchline{}{}{}{}{}
%

\title{The A-Cycle Problem In XY model with Ring Frustration}

\author{Jian-Jun Dong}
\author{Peng Li}

\address{College of Physical Science and Technology, Sichuan University, 610064,
Chengdu, China}
\address{Key Laboratory of High Energy Density Physics and Technology of Ministry of
Education, Sichuan University, Chengdu, 610064, China
\\
lipeng@scu.edu.cn}

\maketitle

\begin{history}
\received{(Day Month Year)}
\revised{(Day Month Year)}
\end{history}

\begin{abstract}
Traditionally, the transverse spin-1/2 XY model is mapped to a fermionic "c-cycle" problem,
where the prior periodic boundary condition is applied to the fermionic chain
and the additional boundary term has been neglected. However, the "a-cycle" problem
(the original problem without any approximation) has not been treated seriously up to now.
In this paper, we consider the XY model with ring frustration and
diagonalize it without any approximation with the help of parity constraint.
Then two peculiar gapless phases have been found.
\end{abstract}


\section{Introduction}

The spin-1/2 XY model in a transverse magnetic filed%
\begin{equation}
H=\sum_{j=1}^{N}\left(  \frac{1+\gamma}{2}\sigma_{j}^{x}\sigma_{j+1}^{x}%
+\frac{1-\gamma}{2}\sigma_{j}^{y}\sigma_{j+1}^{y}\right)  -h\sum_{j=1}%
^{N}\sigma_{j}^{z}, \label{HXY}%
\end{equation}
with Pauli matrices $\sigma_{j}^{\alpha}$ ($\alpha$=$x,y,z$) and ring
structure $\sigma_{N+1}^{\alpha}$=$\sigma_{1}^{\alpha}$ is one of the most
fundamental systems endowed with interesting phenomenon, which plays a
central role in the study of many-body quantum systems
\cite{Lieb,Katsura,McCoy}. It can be mapped to a model of spinless fermions
using the Jordan-Wigner transformation $\sigma_{j}^{+}$=$\left(  \sigma
_{j}^{x}+\operatorname{i}\sigma_{j}^{y}\right)  /2$=$c_{j}^{\dag}\exp\left(
\operatorname{i}\pi\sum_{l<j}c_{l}^{\dag}c_{l}\right)  $, which map an
$\uparrow$ spin or a $\downarrow$ spin at any site to the presence or absence
of a spinless fermion at that site \cite{J-W}. Following the Jordan-Wigner
transformation, Eq. (\ref{HXY}) takes the form%
\begin{align}
H  &  =Nh-2h\sum_{j=1}^{N}c_{j}^{\dag}c_{j}+\sum_{j=1}^{N-1}\left(  \gamma
c_{j}^{\dag}c_{j+1}^{\dag}+c_{j}^{\dag}c_{j+1}+h.c.\right) \nonumber\\
&\quad  -\exp(\operatorname{i}\pi M)\left(  \gamma c_{N}^{\dag}c_{1}^{\dag}%
+c_{N}^{\dag}c_{1}+h.c.\right)  , \label{HXY fermion}%
\end{align}
where $M=\sum_{j=1}^{N}c_{j}^{\dag}c_{j}$ is the total number of fermions.
Traditionally, one can get a standard quadratic Hamiltonian which can be
diagonalized easily after applying periodic boundary condition (PBC) to the
fermionic chain and neglecting the additional term $\left[  \exp
(\operatorname{i}\pi M)+1\right]  \left(  \gamma c_{N}^{\dag}c_{1}^{\dag
}+c_{N}^{\dag}c_{1}+h.c.\right)  $ in the thermodynamic limit
\cite{Pfeuty,Suzuki}. This is the "c-cycle" problem for the XY model
\cite{Lieb}.

Recently, we have shown that for the odd-numbered antiferromagnetic Ising Model
$\left(  \gamma=1\right)  $, i.e. the Ising Model which suffers a ring
frustration  \cite{Dong}, the "c-cycle" problem is not equivalent to the
original periodic Ising chain even in the thermodynamic limit so we must
consider the "a-cycle" problem, i.e. the original problem without any
approximation. In fact, we impose PBC on the original spin model rather than
on the fermionic chain and we have no reason to neglect the additional term.
In this sense, the "a-cycle" problem is more appropriate. In this paper, we
will focus on the "a-cycle" problem of an periodic antiferromagnetic XY chain.

The paper is organized as follows: In Section \ref{Diagonalisation}, we
diagonalize the XY model with ring frustration for the "a-cycle" problem. In
Section \ref{Phase diagram}, we analyse the phase diagram. Finally, we draw our
conclusion in Section \ref{Conclusion}

\section{Diagonalization}

\label{Diagonalisation}

We only consider a periodic chain with an odd number of lattice sites because
it suffers a ring frustration \cite{Dong}. In the limit $h\rightarrow0$, the fully
anti-aligned Ne\'{e}l state cannot complete periodically, as it requires an
even total number of spins \cite{Solomon,Owerre}.

Eq. (\ref{HXY fermion}) is not a standard quadratic Hamiltonian due to the
presence of the factor $\exp(\operatorname{i}\pi M)$. To diagonalize this
Hamiltonian without any approximation, we should deal with it carefully.
First, we note that though the total number of fermions $M$ does not conserve,
its evenness or oddness is invariant, so that the parity of the system
$P$=$\exp\left(  \text{i}\pi M\right)  $ is invariant \cite{Lieb}. When $M\in
even$, $P=1$, we will refer to as even channel, we can define anti-PBC
condition $c_{1}^{\dag}=-c_{N+1}^{\dag}$. When $M\in odd$, $P=-1$, we will
refer to as odd channel, and we can define PBC condition $c_{1}^{\dag}%
=c_{N+1}^{\dag}$. In both cases, Eq. (\ref{HXY fermion}) can be rewritten as%
\begin{equation}
H=Nh-2h\sum_{j=1}^{N}c_{j}^{\dag}c_{j}+\sum_{j=1}^{N}\left(  \gamma
c_{j}^{\dag}c_{j+1}^{\dag}+c_{j}^{\dag}c_{j+1}+h.c.\right)  \label{HXY2}%
\end{equation}
The Hamiltonian Eq. (\ref{HXY2}) is quadratic form in fermion creation and
annihilation operators and can be diagonalized by introducing Fourier
transformation $c_{q}$=$\frac{1}{\sqrt{N}}\sum_{j=1}^{N}c_{j}\exp\left(
\operatorname{i}qj\right)  $ and Bogoliubov transformation $\eta_{q}$%
=$u_{q}c_{q}-$i$v_{q}c_{-q}^{\dagger}$, where the lattice spacing $a$ is set
to be unit and the momentum $q$ lies in the first Brillouin zone $\left(
-\pi,\pi\right]  $ and is quantized in units of $2\pi/N$ \cite{Sachdev,Dutta}.
The even channel implies the momentum $q$ must take a value in the set%
\begin{equation}
q^{e}=\{-\frac{N-2}{N}\pi,\ldots,-\frac{1}{N}\pi,\frac{1}{N}\pi,\ldots
,\frac{N-2}{N}\pi,\pi\},
\end{equation}
while the odd channel means the momentum $q$ must take a value in the
following set%
\begin{equation}
q^{o}=\{-\frac{N-1}{N}\pi,\ldots,-\frac{2}{N}\pi,0,\frac{2}{N}\pi,\ldots
,\frac{N-1}{N}\pi\},
\end{equation}
where the superscript $e$ denotes even channel and $o$ denotes odd channel
\cite{Dong}.

We can arrive at the diagonalized Hamiltonian given by%
\begin{equation}
H^{e}=2\sum_{q\in q^{e},q\neq\pi}\omega(q)\eta_{q}^{\dagger}\eta_{q}%
+\Lambda_{e}+2\left(  h+1\right)  -2\left(  h+1\right)  c_{\pi}^{\dagger
}c_{\pi}. \label{He}%
\end{equation}%
\begin{equation}
H^{o}   =2\sum_{q\in q^{o},q\neq0}\omega(q)\eta_{q}^{\dagger}\eta
_{q}+\Lambda_{o}+\left\vert h-1\right\vert +\left(  h-1\right)
-2\left(  h-1\right)  c_{0}^{\dagger}c_{0}. \label{Ho}
\end{equation}
with%
\begin{align}
\omega(q)  &  =\sqrt{\left(  \cos{q}-h\right)  ^{2}+\left(  \gamma\sin
{q}\right)  ^{2}},2u_{q}v_{q}=\frac{\gamma\sin{q}}{\omega(q)}\\
u_{q}^{2}  &  =\frac{\omega(q)+\cos{q}-h}{2\omega(q)},v_{q}^{2}=\frac
{\omega(q)-\cos{q}+h}{2\omega(q)}\\
\Lambda_{e}  &  =-\sum_{q^{e}}\omega(q),\Lambda_{o}=-\sum_{q^{o}}\omega(q)
\end{align}
We must notice that there is no need of Bogoliubov transformation for $q=\pi$
and $q=0$. The Hamiltonian $H^{e}$ means only even number fermionic occupation
states are valid states for the original spin model due to the even parity
constraint, such as $|\phi^{e}\rangle$, $\eta_{\frac{\pi}{N}}^{\dagger}c_{\pi
}^{\dagger}|\phi^{e}\rangle$, $\eta_{\frac{\pi}{N}}^{\dagger}\eta_{\frac{3\pi
}{N}}^{\dagger}|\phi^{e}\rangle$, where the vacuum state $|\phi^{e}\rangle$ is
defined as $c_{\pi}|\phi^{e}\rangle=0$, $\eta_{q}|\phi^{e}\rangle=0$ for all
$\left\{  q\in q^{e}|q\neq\pi\right\}  $. While the Hamiltonian $H^{o}$
implies that only odd number fermionic occupation states are correct states,
such as $c_{0}^{\dagger}|\phi^{o}\rangle$, $\eta_{\frac{2\pi}{N}}^{\dagger
}|\phi^{o}\rangle$, $\eta_{\frac{2\pi}{N}}^{\dagger}\eta_{\frac{-2\pi}{N}%
}^{\dagger}c_{0}^{\dagger}|\phi^{o}\rangle$. Similarly, the vacuum
state\ $|\phi^{o}\rangle$ satisfies $c_{0}|\phi^{o}\rangle=0$, $\eta_{q}%
|\phi^{o}\rangle=0$ for all $\left\{  q\in q^{o}|q\neq0\right\}  $ . Due to
the parity constraint \cite{Schultz}, the valid states for the both channels
are $2^{N-1}$, thus we can get $2^{N}$ states totally which is in accordance
with the original spin model. The core of the "a-cycle" problem lies in this
parity constraint which helps us to obliterate the redundant degree of freedom
in each channel exactly and reconstruct the correct states of the original
spin problem. On the contrary, the parity factor $\exp(\operatorname{i}\pi M)$ in the
"c-cycle" problem is not treated seriously \cite{Lieb}.

\section{The Phase diagram}

\label{Phase diagram}

\begin{figure}[ptb]
\begin{center}
\includegraphics[width=0.4\textwidth]{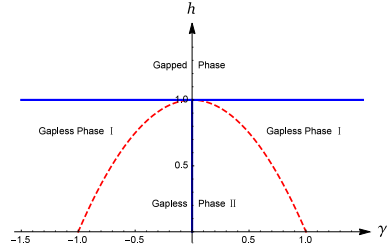}
\end{center}
\caption{The phase diagram of the transverse XY model with ring frustration for the "a-cycle"
problem. The horizontal line $h=1$ denotes the phase transition between the
gapped phase and the gapless phase \uppercase\expandafter{\romannumeral1}. The
vertical line $\gamma=0$, $0<h<1$ stands for the phase transition
in the gapless phase \uppercase\expandafter{\romannumeral2}. The dotted line
given by the equation $\gamma^{2}+h-1=0$, denotes the boundary between the
gapless phase \uppercase\expandafter{\romannumeral1} and
\uppercase\expandafter{\romannumeral2}. The ground state in the gapless phase
\uppercase\expandafter{\romannumeral1} and gapped phase is $c_{0}^{\dagger
}|\phi^{o}\rangle$, the $2$-fold degenerate ground states in the gapless phase
\uppercase\expandafter{\romannumeral2} are given by $\eta_{k^{e}}^{\dagger
}c_{\pi}^{\dagger}|\phi^{e}\rangle$, $\eta_{-k^{e}}^{\dagger}c_{\pi}^{\dagger
}|\phi^{e}\rangle$ or $\eta_{k^{o}}^{\dagger}|\phi^{o}\rangle$, $\eta_{-k^{o}%
}^{\dagger}|\phi^{o}\rangle$, depending on the lattice number $N$ and
parameter $h$, $\gamma$.}%
\label{fig1}%
\end{figure}

In this section we discuss the phase diagram of the transverse XY ring for the
"a-cycle" problem. As shown in Fig. \ref{fig1}, there are two special gapless
phase which is not exist in the "c-cycle" problem. As a function of $q$, the
dispersion relation $\omega(q)$ has a minimum at the point $0$ in the gapped
phase and the gapless phase \uppercase\expandafter{\romannumeral1}. While in
the gapless phase \uppercase\expandafter{\romannumeral2}, the point is $k_{0}$
which is given by $k_{0}=\arccos\left(  \frac{h}{1-\gamma^{2}}\right)  $. In
Ref. \cite{Dong}, the special gapless phase for the Ising case $\left(
\gamma=1,h<1\right)  $ has been discussed in detail by the aid of band
structure analysis method. In the Ising case, when $h=0$, the fully
anti-aligned Ne\'{e}l state cannot complete periodically due to the ring
frustration. Thus there has to be at least one defect, which are sometimes
called kink or domain wall \cite{Owerre,Nagaosa}. The ground state is $2N$
fold degenerate because the position of the up-up or down-down kink is
arbitrary. When a small transverse field is applied, as a source of quantum
tunnelling, the high level of degenerate ground state would split into a
gapless band. Similarly, a transverse XY ring with an odd number of lattice
sites would also possess the special gapless phase. Next, we will discuss the
phase diagram in detail.

In the gapped phase $\left(  h>1\right)  $, the lowest energy states in even
channel are $\eta_{\frac{\pi}{N}}^{\dagger}c_{\pi}^{\dagger}|\phi^{e}\rangle$
and $\eta_{-\frac{\pi}{N}}^{\dagger}c_{\pi}^{\dagger}|\phi^{e}\rangle$ rather
than the vacuum state $|\phi^{e}\rangle$\ due to the presence of the minus
sign in the term $-2\left(  h+1\right)  c_{\pi}^{\dagger}c_{\pi}$ in Eq.
(\ref{He}). The corresponding lowest energy is given by $2\omega(\frac{\pi}%
{N})+\Lambda_{e}$. Likewise, the lowest energy state in odd channel is
$c_{0}^{\dagger}|\phi^{o}\rangle$, and its energy reads $\Lambda_{o}$. In the
thermodynamic limit, we have $\Lambda_{e}=\Lambda_{o}=-N\int_{-\pi}^{\pi}%
\frac{dq}{2\pi}\omega(q)$. Compare the expression $2\omega(\frac{\pi}%
{N})+\Lambda_{e}$ with $\Lambda_{o}$, we found that the true ground state
comes from odd channel, i.e. $|GS\rangle=c_{0}^{\dagger}|\phi^{o}\rangle$. And
the first excited state comes from even channel. Thus the energy gap in this
region is given by $2\left(  h-1\right)  $ when $N\rightarrow\infty$.

In the gapless phase \uppercase\expandafter{\romannumeral1} $\left(
0<h<1,\gamma^{2}+h-1>0\right)  $, using the same method, we can found that the
ground state is also given by $c_{0}^{\dagger}|\phi^{o}\rangle$. But the
corresponding energy is $\Lambda_{o}+2\left(  1-h\right)  $ rather than
$\Lambda_{o}$, because the absolute value\ of the term $\left\vert
h-1\right\vert $ in Eq. (\ref{Ho}) takes different value in different regions.
We can rewrite the ground state energy as $2\omega(0)+\Lambda_{o}$. Together
with the excited energy $2\omega(\frac{2\pi}{N})+\Lambda_{o}$, $2\omega
(\frac{4\pi}{N})+\Lambda_{o}$, etc. in odd channel and the excited energy
$2\omega(\frac{\pi}{N})+\Lambda_{e}$, $2\omega(\frac{3\pi}{N})+\Lambda_{e}$,
etc. in even channel, they form a continuous gapless energy band in the
thermodynamic limit $N\left(  odd\right)  \rightarrow\infty$. A quantum phase
transition occurs at a value of transverse field given by $h=1$, because the
second derivative of the ground state energy per site $\int_{-\pi}^{\pi}%
\frac{dq}{2\pi}\omega(q)$ has a divergent peak in this line.

\begin{table}[pt]
\tbl{The states $\eta_{k^{o}\pm\frac{2n\pi}{N}}^{\dagger}|\phi^{o}%
\rangle\left(  k^{o}\pm\frac{2n\pi}{N}\in q^{o}\right)  $ in odd channel and
the states $\eta_{k^{e}\pm\frac{2m\pi}{N}}^{\dagger}c_{\pi}^{\dagger}|\phi
^{e}\rangle\left(  k^{e}\pm\frac{2m\pi}{N}\in q^{e}\right)  $ in even channel
form a continuous gapless energy band in the thermodynamic limit ($m$ and $n$
are integer). The energy value of the corresponding state is readily read out
from the diagonalized Hamiltonian, Eq. (\ref{He}) or Eq. (\ref{Ho}).}
{\begin{tabular}
[t]{|l|l|l|l|}\hline
\multicolumn{2}{|c|}{Odd channel} & \multicolumn{2}{|c|}{Even channel} \\
\hline
States & Energy & States & Energy\\\hline
\multicolumn{1}{|c|}{$\eta_{k^{o}}^{\dagger}|\phi^{o}\rangle,\eta_{-k^{o}%
}^{\dagger}|\phi^{o}\rangle$} & \multicolumn{1}{|c|}{$2\omega(k^{o}%
)+\Lambda_{o}$} & \multicolumn{1}{|c|}{$\eta_{k^{e}}^{\dagger}c_{\pi}%
^{\dagger}|\phi^{e}\rangle$,$\eta_{-k^{e}}^{\dagger}c_{\pi}^{\dagger}|\phi
^{e}\rangle$} & \multicolumn{1}{|c|}{$2\omega(k^{e})+\Lambda_{e}$}\\\hline
\multicolumn{1}{|c|}{$\eta_{k^{o}\pm\frac{2\pi}{N}}^{\dagger}|\phi^{o}\rangle
$} & \multicolumn{1}{|c|}{$2\omega(k^{o}\pm\frac{2\pi}{N})+\Lambda_{o}$} &
\multicolumn{1}{|c|}{$\eta_{k^{e}\pm\frac{2\pi}{N}}^{\dagger}c_{\pi}^{\dagger
}|\phi^{e}\rangle$} & \multicolumn{1}{|c|}{$2\omega(k^{e}\pm\frac{2\pi}%
{N})+\Lambda_{e}$}\\\hline
\multicolumn{1}{|c|}{$\eta_{k^{o}\pm\frac{4\pi}{N}}^{\dagger}|\phi^{o}\rangle
$} & \multicolumn{1}{|c|}{$2\omega(k^{o}\pm\frac{4\pi}{N})+\Lambda_{o}$} &
\multicolumn{1}{|c|}{$\eta_{k^{e}\pm\frac{4\pi}{N}}^{\dagger}c_{\pi}^{\dagger
}|\phi^{e}\rangle$} & \multicolumn{1}{|c|}{$2\omega(k^{e}\pm\frac{4\pi}%
{N})+\Lambda_{e}$}\\\hline
\multicolumn{1}{|c|}{$\vdots$} & \multicolumn{1}{|c|}{$\vdots$} &
\multicolumn{1}{|c|}{$\vdots$} & \multicolumn{1}{|c|}{$\vdots$}\\\hline
\end{tabular} }
\end{table}

In the gapless phase \uppercase\expandafter{\romannumeral2} $\left(
0<h<1,\gamma^{2}+h-1<0\right)  $, as mentioned before, $\omega(q)$ has a
minimum at the point $k_{0}$ which is given by $k_{0}=\arccos\left(  \frac
{h}{1-\gamma^{2}}\right)  $. Thus the ground state in this phase is no longer
given by $c_{0}^{\dagger}|\phi^{o}\rangle$. To obtain the properties in the
thermodynamic limit, we can keep $N$ as a variable, and then let
$N\rightarrow\infty$. The special point $k_{0}$ which depends on the
transverse filed $h$ and the anisotropy parameter $\gamma$ may neither belong
to $q^{e}$ nor belong to $q^{o}$ for an finite chain because $q^{e}$ and
$q^{o}$ are quantized in units of $2\pi/N$. So the problem became even more
complicated in this phase. But for an arbitrary finite chain, we can always
find some special values $k^{e}\in q^{e}$ in even channel and $k^{o}\in q^{o}$
in odd channel that minimize the dispersion relation respectively. Then the
$2$-fold degenerate lowest energy states and the corresponding energy in even
channel are $\eta_{k^{e}}^{\dagger}c_{\pi}^{\dagger}|\phi^{e}\rangle$,
$\eta_{-k^{e}}^{\dagger}c_{\pi}^{\dagger}|\phi^{e}\rangle$ and $2\omega
(k^{e})+\Lambda_{e}$. Similarly, in odd channel, they are given by
$\eta_{k^{o}}^{\dagger}|\phi^{o}\rangle$, $\eta_{-k^{o}}^{\dagger}|\phi
^{o}\rangle$ and $2\omega(k^{o})+\Lambda_{o}$, respectively. The ground states
can come from even channel or odd channel, depending on the lattice number $N$
and parameter $h$, $\gamma$. One can check that $k^{e}\rightarrow k_{0}$ and
$k^{o}\rightarrow k_{0}$ as the number of lattice sites $N\rightarrow\infty$.
Thus in the thermodynamic limit, the ground state energy per site is given by
$2\omega(k_{0})/N-\int_{-\pi}^{\pi}\frac{dq}{2\pi}\omega(q)$. Its second
derivative has a divergent peak in the line $\gamma=0$, $0<h<1$. The system
undergoes a quantum phase transition in this line. Furthermore, to understand
the formation of the gapless energy band, we also list some excited states in
Table 1.

\section{Conclusion}

\label{Conclusion}

In this paper we have provided an extensive discussion of the "a-cycle"
problem of an odd-numbered periodic antiferromagnetic XY chain, i.e. the
transverse XY model with ring frustration. The presented material
covers several aspects. First, we pay a special attention to the parity of the
system $P$=$\exp\left(  \text{i}\pi M\right)  $ which plays an important role
in the "a-cycle" problem. We can diagonalize the original Hamiltonian in two
channels, depending on the parity of the system. Secondly, with the help of the
parity constraint, the true ground states in different phases are obtained.
When the applied transverse field is weak $h<1$, two peculiar gapless phases
are founded due to the ring frustration.

We acknowledge useful discussions with Yan He. This work was supported by the
NSFC under Grants no. 11074177, SRF for ROCS SEM (20111139-10-2).

\end{document}